\documentclass[a4paper]{article}

\usepackage[english]{babel}
\usepackage[T1]{fontenc}

\usepackage[a4paper,top=3cm,bottom=2cm,left=3cm,right=3cm,marginparwidth=1.75cm]{geometry}

\usepackage{amsmath}
\usepackage{graphicx}
\usepackage[colorinlistoftodos]{todonotes}
\usepackage[colorlinks=true, allcolors=blue]{hyperref}

\title{Counter-Geoengineering: Feasibility and Policy Implications for a Geoengineered World}
\author{Felipe de Bolle and Egemen Kolemen}

\begin{document}
\maketitle

\begin{abstract}
With the increasing urgency of climate change’s impacts and limited success in reducing emissions, “geoengineering,” or the artificial manipulation of the climate to reduce warming rates, has been proposed as an alternative short-term solution. Options range from taking carbon out of the atmosphere through carbon sinks and brightening clouds to increasing the planet’s albedo through the release of reflective particles into the atmosphere. While still controversial, geoengineering has been proposed by some as a promising and low-cost way of combating climate change. In particular, so-called ‘moderate’ geoengineering is claimed to be achievable with few potential side effects or other ramifications. However, this paper argues that the effect of moderate geoengineering can easily be nullified by ‘counter-geoengineering,’ and any impactful geoengineering would require a global governance framework to prevent countries which benefit from warming temperatures from deploying counter-geoengineering. In this paper, we take Russia as an example due to its potential interest in counteracting geoengineering and its significant ability to release a great amount of methane, a viable counter-geoengineering pathway in the short term. 
  
\end{abstract}

\section*{Introduction}

Various agreements to curb emissions in an effort to combat climate change have been made in the last several decades. The increasing urgency of climate change has led much of the world to plan for a transition to a “green” economy that does not release high amounts of greenhouse gasses. This transition, however, is costly and takes time. By the time the world is fully carbon neutral, global temperatures could be several degrees higher than in pre-industrial times, having disastrous consequences for the world’s ecosystems and our capacity to live comfortably. As a result, solutions like geoengineering have also been discussed as a way to stop, or even reverse, the rise in global temperatures without having to completely halt carbon emissions. Although proponents highlight many potential benefits, we argue that geoengineering’s promise rests on the unrealistic assumption of a high level of global agreement and cooperation without retaliation. The warming of the world’s temperatures, although destructive to ecosystems worldwide and deadly in the Global South, has led to benefits such as a decrease in climate-related deaths and a longer growing season in some regions, particularly at higher latitudes. 

Geoengineering most likely will have negative implications for some countries. Here, we look at the case study of Russia, a notable geopolitical actor. Russia, as the largest country in the Arctic Region, may oppose methods that aim to reduce world temperatures by artificial means for various reasons, one being its increased access to shipping routes through the Arctic Ocean. The country thus serves as a good example of a world power with a strong incentive to prevent initiatives that would artificially reduce global temperatures. Russia, as one of the world’s leading producers of natural gas, has significant influence over the global response to climate change and considerable capacity to counteract geoengineering solutions. Natural gas is one of the leading causes of methane emissions, which in the short-term can have up to 120 times the warming capacity of carbon dioxide. To counteract the effects of moderate geoengineering solutions in the first few years, a country like Russia could relatively easily do so at low cost and in a short time by increasing its leakage rate from natural gas production and transportation to around 25\%. Escalating geoengineering deployment from moderate to large deployment is an option to counteract increased methane. However, this approach could lead to significantly increased side effects, including altered weather patterns and disturbed ecosystems, and undermine the incremental, “realistic” geoengineering approach many in the community advocate for. Ultimately, the possibility of counter-geoengineering renders geoengineering deployment problematic due to governance issues. The versions of geoengineering that may be moderate enough as to not cause substantial complexities are susceptible to being counteracted. Consequently, despite the presumed technical feasibility and cost-effectiveness of geoengineering, governance challenges that result from the threat of counter-geoengineering overshadow these benefits, making a strategy focused on reducing or achieving negative greenhouse gas emissions a more viable approach for mitigating climate change.

\section*{Literature Review}

Geoengineering, or technological strategies to combat climate change by modifying the climate without curbing carbon emissions, has been a topic of discussion for several decades, but it has only reached the mainstream of climate change policy debates relatively recently (Möller 2023). A variety of geoengineering solutions have been proposed, broadly falling under two broad categories: solar radiation management (SRM) and carbon dioxide removal. SRM techniques include placing “white roofs” over urban areas or deserts to increase the amount of sunlight reflected back to space, whitening clouds, and releasing stratospheric aerosols, usually sulfate aerosols, to increase Earth’s albedo. Carbon dioxide removal strategies, on the other hand, might involve implementing wide reforestation, sequestering biomass and biochar, enhancing weathering, capturing carbon dioxide from ambient air, or storing carbon dioxide in the deep ocean, where it can stay for hundreds of years, through ocean fertilization (UK Royal Society 2009). As none of these solutions have been implemented, and it is extremely difficult to test them ethically on a wide scale, there is much uncertainty about the impacts of geoengineering as well as potential adverse effects. As a result, estimates for the impact of the various proposed solutions in curbing warming vary substantially across studies. This paper particularly focuses its analysis on sulfate aerosol SRM as it is the most often studied form of geoengineering and the one with the most promise, especially in regard to its potential low cost and high effectiveness. However, the paper’s conclusions can be extended to other forms of modifying the climate on a global scale.

The idea of global climate modification started being proposed in the 1990s as a more economically realistic and politically agreeable alternative to reducing greenhouse gas emissions. As argued by Thomas Schelling in 1996, geoengineering would not have to contend with large-scale behavioral change, discrepancies in national regulations, or a universal commitment to reducing emissions. Instead, it would simply involve a decision of “what to do, how much to do, and who to pay for it” (Schelling, 1996: 303). The last of these three points–geoengineering’s cost–has particularly been hailed as a significant advantage it has over other methods of combating climate change. Recent studies have estimated that geoengineering would be considerably less costly from an economic standpoint than changing the world’s energy structure in a rapid timeframe. This appears especially significant in the case of SRM, which would entail releasing large amounts of hydrogen sulfide or sulfur dioxide into the stratosphere to increase the amount of sunlight reflected. The cost of deploying and maintaining sulfate aerosols at a large scale into the atmosphere–the most commonly proposed process to enact solar geoengineering–is estimated by Smith and Wagner (2018) at around \$2.25 billion per year, with \$3.5 billion in total pre-start costs. Estimates vary, but this estimated cost is in line with other studies that have addressed this question (Barrett 2008, Mahajan \textit{et al}. 2018). By contrast, the International Energy Agency estimated in 2023 that the total cost of going “net zero” globally stands at \$4 trillion per year over the next 30 years, resulting in a total cost “on the order of \$100 trillion to \$120 trillion, plus or minus a few trillion here and there” (Crawford 2023). Taking these estimates, SRM seems to be over a thousand times less costly than drastically reducing carbon emissions. As previously mentioned, all estimates regarding geoengineering are subject to substantial variation due to their speculative nature, but even the most pessimistic estimates of the economic cost of implementing SRM are markedly lower than the presumed cost of going net zero. Other geoengineering methods have greatly varying estimated costs, though many remain significantly under the estimated figure for changing the world’s energy structure. The costlier alternatives include sequestering carbon dioxide from the air and increasing surface albedo whereas other low-cost solutions include cloud whitening and some forms of ocean fertilization. Overall, the cost effectiveness of SRM (or the amount of radiative forcing per amount of money spent) seems to outweigh that of most other geoengineering solutions, explaining the increasing academic focus on this specific form of geoengineering (UK Royal Society 2009: 20, 35).

Turning more specifically to SRM, there has been a fair amount of research on its potential impact, though estimates again vary. As the cost of SRM is estimated to be very low, the scale of sulfate aerosol release will largely depend on risk assessments rather than economic restraints, which causes a wide range of potential responses. Additionally, SRM involves considerations of where to release aerosols, how high in the atmosphere to do so, and what particle size to use, all of which contribute to more noise in potential impact estimates. Finally, as analyzed by Lunt \textit{et al}. (2008), the effects of solar geoengineering may involve a considerable amount of “spatial heterogeneity,” meaning that different areas of the globe may be impacted differently, such as the tropics cooling more than high latitudes. Overall, in order to reverse a doubling of pre-industrial carbon dioxide, with a current radiative forcing of 4 W/m\textsuperscript{2}, around 1.7\% of the sunlight reaching Earth would have to be reflected by sulfate aerosols (Caldeira \& Wood 2008, Govindasamy \& Caldeira 2000, Govindasamy \textit{et al.} 2002, Lunt \textit{et al.} 2008), though it is difficult to obtain very precise numbers as a result of the uncertainty of the response of climate feedback systems. Some models for stratospheric sulfate aerosol impact such as Rasch \textit{et al}. (2008a) and Robock \textit{et al}. (2008) have considered the effect of different sizes and deployment strategies of aerosols, concluding that impacts may vary quite significantly depending on these considerations. Nevertheless, most model results suggest that stratospheric aerosol geoengineering could have strong impacts on global temperatures in a short timeframe. Depending on the scale and specificities of the geoengineering solution implemented, global temperatures could drop to pre-industrial levels within a decade (Matthews and Caldeira 2007). However, it has also been shown by multiple of these same studies that a sudden reversal of geoengineering policies in a high-carbon environment, or a breakdown or failure in the scheme, could lead to an extreme level of sudden atmospheric warming. A geoengineered world with low temperatures and increasing carbon emissions would mean a greater level of carbon stored in the surface ocean and land. A sudden increase in incoming solar radiation would not only warm the world through the direct effect of lower global albedo, but it would also trigger further warming through the release of stored carbon. The consequences of this sudden warming could be greatly worse than the current effects of global warming, which has seen temperatures rise relatively steadily and over relatively long periods of time, at least in comparison to what would happen in a suddenly un-geoengineered world. Matthews and Caldeira (2007) suggest that warming rates from a sudden reversal or failure of geoengineering could be up to 20 times greater than present-day rates, which could be catastrophic for the global climate system. Therefore, SRM–although potentially very effective in short timescales and less costly than reaching net zero or implementing most other methods of geoengineering–also poses severe risks for the global climate.

The risks associated with a sudden breakdown in stratospheric aerosol geoengineering, as well as other ethical considerations such as unequal impact in different areas of the world and the potential discouragement of tackling the issue at its source by reducing greenhouse gas emissions, has led to a great number of critiques of SRM in recent years. Stratospheric aerosol geoengineering’s side effects have not been extensively researched, and the exact risks posed by it are not fully understood. Besides the potential risk of extremely rapid warming following an interruption in the technology, potential effects on the hydrological cycle, adverse effect on ozone layer recovery (Tilmes \textit{et al}. 2008), increased stratosphere-troposphere exchange which could effect high-altitude tropospheric clouds (Joshi and Shine 2003), and possible changes in biological productivity as a result of potential changes in the carbon cycle (IPCC 2007). In February of 2024, Switzerland’s proposal for a UN expert group on solar geoengineering was dismissed by a United Nations summit after no consensus was reached at the UN Environment Assembly (Civillini 2024 and Limb 2024). The dismissal of a research group on geoengineering, which would only be the first step for a potential geoengineered future, underlines how contentious the topic has become recently. In such an environment, and with such risks associated with a high level of geoengineering, many advocates of the technology have sought to propose more “moderate” forms of mitigating global warming through stratospheric aerosols. 

For the analysis of this paper, a solution based on Keith and MacMartin (2015) will be considered. In their paper, Dr. David Keith and Dr. Douglas MacMartin seek to address many of the common concerns associated with geoengineering by proposing an implementation that they claim to be “temporary, moderate and responsive.” They posit that most negative discourse around geoengineering’s risks results from assumptions about the level of geoengineering implemented, and that with a more moderate version of geoengineering there would be more moderate or insignificant potential side effects for the world’s ecosystems. Their solution does not seek to fully restore global temperatures to pre-industrial times, for example, but rather only to offset around half of the growth of human-caused increase in temperature. Moreover, it is a plan that has progressively diminishing solar geoengineering to zero as an end goal, which alleviates worries about the potential impacts of a sudden reversal of geoengineering policies. As such, it is a solution that might prompt closer investigation. The numbers used in this paper for the amount of stratospheric sulfur aerosols released into the atmosphere are based on the “specific scenario” shown in Figure 1 of Keith and MacMartin’s paper. They note, citing Pierce \textit{et al} (2010), that for radiative forcing of less than 0.5 W/m\textsuperscript{2} the radiative forcing efficacy of 1 million tons of sulfur (MtS) for most proposed methods of introducing sulfate into the atmosphere stands around 0.6-0.8 W/m\textsuperscript{2}. In this paper, we consider the higher end of this range for our estimates. Ultimately, Keith and MacMartin estimate that a potential scenario that implements a “temporary, moderate, and responsive” scenario of solar geoengineering might include an initial release of 0.035 MtS in the first year, with the injection rate rising by 0.035 MtS/yr in following years so that after a decade it stands at 0.35 MtS/yr. These are the numbers used in the calculations of this paper.

The solution is crafted using a central planner model and only claims to be “less suboptimal” than other proposed implementations of geoengineering, also acknowledging that they do not go into detail into potential social or political implications of SRM. The question that this paper seeks to address is exactly whether this moderate solution would be feasible given the many social and political actors with a vested interest in continuing climate change. Although a moderate version of geoengineering would not cause as extreme shifts in the trajectory of global temperatures as many other proposed solutions would, it would still provide a strong pushback against the current rate of warming. As this paper argues, some countries might have an incentive to preserve the current rate of warming, as the effects of global warming are not entirely harmful across the board for everyone. In the context of solar geoengineering, this could lead to a sort of “counter-geoengineering,” a topic previously investigated in the literature by Biermann \textit{et al}. (2022), Heyen \textit{et al}. (2019), and Parker \textit{et al}. (2018). We seek to fill the gap in the existing literature by showing that “moderate” solutions to geoengineering are highly susceptible to successful counter-geoengineering by governments or entities that might have an incentive to maintain our current rate of warming or prevent a reversal in the trajectory of global temperatures. 

\section*{The Impacts of Climate Change}

Climate change and its many impacts are undoubtedly one of the greatest challenges that our planet currently faces. Combating the effects of climate change and reducing emissions to reduce our impact on the climate has been the topic of several international discussions and agreements in the last couple of decades. It is clear that climate change has already had some catastrophic consequences, such as coastal inundation, intensified natural disasters, and loss of biodiversity, all of which will continue to intensify in the coming decades if the world keeps following the current trajectory. However, even with the many indisputably adverse effects of rising temperatures, some countries may stand to benefit in some ways from rising temperatures. The heterogeneous impacts of increased temperatures may pose a challenge to a strong unified global response to climate change. In the case of geoengineering, it may incentivize some countries to deviate from a global agreement and implement “counter-geoengineering.”

\subsection*{The Winners from Global Warming}

As with virtually any change, there are positive and negative effects of climate change. Although climate change tends to be discussed in an overwhelmingly negative context, there are some countries that nonetheless stand to benefit, at least in the short term. Particularly in higher latitudes, the increase in global temperatures promises to bring about a more moderate climate, which may result in longer growing seasons and greater access to year-round warm water ports. According to a 2020 study by Deloitte, although most countries will have their economies negatively impacted by climate change, up to 70 countries around the world could get an economic boost from warming temperatures over the following century, with the majority of the benefits concentrated in countries with cold-weather climates (consultancy.eu 2020). This is primarily driven by longer growing seasons and greater levels of agricultural production expected with higher average temperatures. Of course, there are a number of effects of climate change that may also negatively impact agricultural productivity in cold-weather regions, but these would likely be outweighed in the short term. 

Countries farther away from the equator stand to benefit the most from increasing temperatures, with nations like Canada, Russia, Finland, Norway, and Sweden all expected to get an economic boost from global warming. In this paper, the case of Russia in particular will be discussed. Preserving the warming trajectory of global temperatures may be of particular interest to Russia for several reasons. Although most developed countries now have birth rates that are lower than replacement levels, Russia has suffered from especially severe demographic challenges that look to intensify in the next several decades. With an increasing aging population, the deadly effects of extreme cold may be of particular concern. Older people are significantly more at risk from temperature-related deaths, and especially cold-related deaths (Chen \textit{et al}. 2024). The mortality burden for cold-related deaths, which happen primarily at high latitudes, is expected to increase as populations age, but global warming may impede much of that increase from happening. Between 2000 and 2019, it is estimated that almost 5 million deaths were associated with temperature extremes, and those associated with cold were 9 times more numerous than those associated with extreme heat (Zhao \textit{et al}. 2021). From 2000-2003 to 2016-2019, the number of cold-related deaths fell considerably, even as some heat-related deaths rose. Overall, the number of people dying from temperature-related causes decreased. It is estimated that Russia’s temperature-related mortality is projected to decrease by 89 deaths per 100,000 people by the end of the century, among the greatest decreases out of any country in the world (Stevens 2023). In a geoengineered world, this decrease in death rate would be significantly lower, and depending on the amount of geoengineering undertaken, Russia could see even more cold-related deaths in the future. A decrease of 89 deaths per 100,000 people by the end of the century, considering Russia’s population of around 144 million, would constitute a decrease of around 128,000 deaths. If it is assumed that “moderate” geoengineering could decrease the rate of warming in half, one could see geoengineering as indirectly killing Russian citizens. Every human interaction that attempts to reduce the global temperature by 1°C could lead to tens of thousands more cold-related deaths in Russia within a decade. If the world undertakes aggressive geoengineering, lowering global temperatures, Russia would likely see an increase in the number of temperature-related deaths on a massive scale. These tens of thousands of additional cold-related deaths from geoengineering could be construed by Russia as an attack on its population and thus give it a reasonable self-defense argument for undertaking counter-geoengineering. In a country which faces an aging population and the burden of demographic decline, minimizing temperature-related mortality is especially important. 

In addition to lessening the demographic issue, global warming has also expanded Russia’s growing season and access to ports in Arctic regions. Climate change is predicted to “positively affect” agriculture in Russia by expanding the growing season in the central and northern regions of the country, even as the southern region may be negatively impacted by a drier climate (Kiselev \textit{et al}. 2013). Perhaps climate change’s biggest benefit for Russia is a warming Arctic Ocean, which is a currently untapped route for shipping due to frozen waters for large parts of the year. With rising temperatures, the Arctic may become a year-round shipping route, which provides great commercial promise for Russia. Additionally, it is estimated that up to 16\% of the world’s oil and 30\% of the world’s natural gas is currently lying underneath the Arctic Ocean, a vast portion of which is under Russian control, and an ice-free Arctic would provide much easier access to the untapped wealth of the ocean (Gatopoulos 2022). Potential benefits of global warming are not exclusive to Russia, but it is perhaps the country with the greatest capacity to counteract potential geoengineering solutions.

\subsection*{Counteracting Geoengineering}

There are many different strategies that a country looking to counteract geoengineering can take to keep global temperatures on an upward trajectory. Heyen \textit{et al}. (2019) explores some of these different approaches to countering geoengineering, focusing its analysis on their implications for the implementation of solar geoengineering, ultimately arguing that the possibility of counter-geoengineering arising from differing temperature preferences may help reduce the “free-driver” problem associated with the unilateral deployment of stratospheric aerosols. The different strategies to enact counter-geoengineering are extensively detailed in Parker \textit{et al}. (2018), which also provides the basis for the analysis of Heyen \textit{et al }(2019). Two different broad approaches to counter-geoengineering exist: “countervailing” approaches, which seek to introduce warming agents into the atmosphere, and “neutralizing” approaches, which seek to remove the SRM agents from the atmosphere or remove their cooling effects. Potential neutralizing approaches may include adding a “base” to the stratosphere that could counteract sulfate aerosols and potentially reduce SRM’s radiative forcing or trailing aircraft deploying stratospheric aerosols and releasing neutralizing agents into the “near-field high-concentration plume” (Parker \textit{et al}. 2018). Another such approach could be the release of aerosols that could deplete black carbon and titanium oxide in the atmosphere, as both substances are cooling agents. Countervailing approaches, which are much more commonly discussed, include the release of greenhouse gasses and stratospheric heating molecules such as sulfur hexafluoride and many different chlorofluorocarbons and hydrochlorofluorocarbons. Potential candidates also include sulfuryl fluoride, HFC-152a, and difluoromethane. Many of these greenhouse gases have very high global warming potentials, meaning a small release could be enough to counteract geoengineering. It is noted in Parker \textit{et al}. (2018), however, that there are not many plausible greenhouse gasses for counter-geoengineering as those with strong global warming potentials tend to have long lifetimes, making them less appealing as agents of countering geoengineering. This is because their impacts would last for much longer than those of sulfur aerosols, which tend to have lifetimes in the scale of several years as opposed to several centuries or millennia. Alternatively, solid particles with high radiative efficiency are also potential candidates for counter-geoengineering, as they tend to have higher radiative forcing per unit mass and shorter atmospheric lifetimes, though their cost at the moment may be prohibitive. 

 In this paper, we focus on methane in our discussion of potential counter-geoengineering, as countries already have the capacity to release large amounts of methane into the atmosphere, it has a relatively short lifetime (around 12 years in the atmosphere, as opposed to around 3,000 years for sulfur hexafluoride, for example), and generally has no negative health effects. Crucially, large amounts of methane can be released in a very short time at a low cost and start warming the atmosphere immediately. The stockpile of methane that already exists makes it a virtually unpredictable agent of counter-geoengineering. Although methane is not as powerful a warming agent as many of the other options discussed in Parker \textit{et al.} (2018), it can have strong warming effects in the short-term. In the first year after deployment, its warming capacity is up to 120 times carbon dioxide’s (Lawton 2021). Its warming capacity does exponentially decrease in the years following deployment, but a constant high rate of methane release could be maintained for a short, sustained period of time so as to successfully counteract “moderate” solar geoengineering. In the more specific application of counter-geoengineering to the example of Russia, methane could be of particular interest as Russia is the world’s second largest producer of methane, not far behind the United States (International Energy Agency 2024).
 
\section*{Calculations}

Taking the example of Russia and methane’s global warming capacity as well as data on methane release from natural gas, one can roughly estimate the amount of additional methane release required for geoengineering. Methane is a very potent greenhouse gas in the short run, which makes its deliberate leakage a useful method for countries looking to geoengineer. Over a 100 year period, methane has a warming capacity of around 30 times carbon dioxide’s, 84 times in a 20-year time scale, and 120 in a 1-year period (Lawton 2021). In calculating the effect of methane leakage, we compare its radiative forcing to that of carbon dioxide, which is estimated at around 1.32 × 10\textsuperscript{-4} W/m\textsuperscript{2} for every part per billion (ppb) in the atmosphere, obtained from the estimated radiative forcing of 3.7 W/m\textsuperscript{2} from a doubling of CO\textsubscript{2} from pre-industrial levels of 280 parts per million (ppm) (IPCC 2021). One part per billion of CO\textsubscript{2} is roughly equivalent to 7.826 × 10\textsuperscript{-3} Gt of carbon dioxide. Using the estimated global warming potential (GWP) of 120 in the first year, compared to a value of 1 for CO\textsubscript{2}, methane’s radiative forcing in the first year after release would be around 0.0016 W/m\textsuperscript{2} for every 7.826 × 10\textsuperscript{-3} Gt of methane.

The estimated radiative forcing of solar geoengineering significantly depends on the specificities of a given geoengineering solution, but it is generally estimated at 0.6-0.8 W/m\textsuperscript{2} for one million tons of sulfur (Pierce \textit{et al}. 2010). Considering the higher end of those estimates and Keith and MacMartin’s scenario of an injection rate of 0.035 MtS per year, with a subsequent increase of 0.035 MtS/yr in each following year (Fig. 1), the radiative forcing of a moderate solar geoengineering scenario would be 0.0245 W/m\textsuperscript{2} in the first year, with a rough linear increase by 0.0245 W/m\textsuperscript{2} per year in the following few years so as to halve the current yearly increase in radiative forcing. Using methane’s estimated radiative forcing, an additional increase of roughly 0.12 Gt of methane in the atmosphere would counteract the first year’s injection rate of 0.035 MtS, which would roughly triple the current yearly rate of increase in atmospheric methane (NOAA 2023). Methane’s potency declines over time, but it remains much stronger than carbon dioxide for the first several years after initial release (Lawton 2021). This means that releasing a constant additional 0.12 Gt of 
\begin{figure}
    \centering
    \includegraphics[width=1\linewidth]{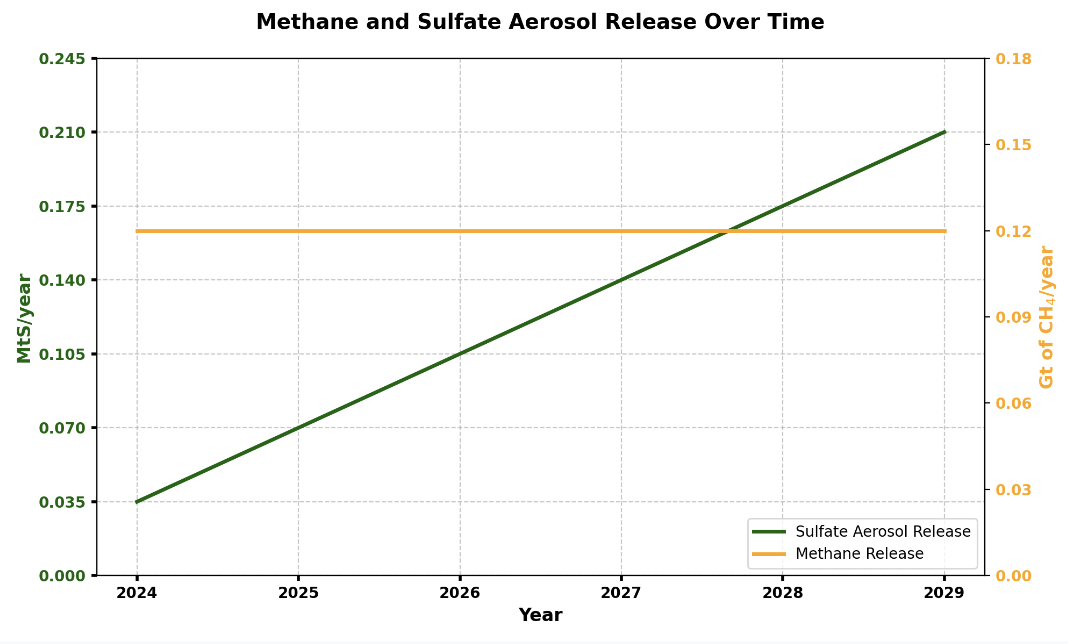}
    \caption{This figure clarifies the amount of methane and sulfur released each year under a possible combination of moderate geoengineering with methane counter-geoengineering.}
    \label{Figure 1}
\end{figure}
methane per year would be able to efficiently counteract geoengineering for the first several years of geoengineering.

One can compare this number to the amount of methane in natural gas production every year. Although it varies year to year, the world produced around 4.4 trillion cubic meters of natural gas in 2022, 0.7 of which were produced in Russia (Enerdata 2024). Assuming that 95\% of natural gas is methane, around 0.665 trillion cubic meters of methane were produced in Russia, or around 0.4767 Gt. To reach the required 0.12 Gt of methane to counter the first year of geoengineering, close to an additional 25\% of Russia’s natural gas production would have to be leaked. Even with a more conservative assumption of 90\% methane in natural gas production, less than an additional 27\% of Russian natural gas would need to be leaked in the first year. Globally, there is wide variation in methane leakage rates from natural gas production, with rates generally averaging 1-3\% of total production (IEA 2024). In Russia, this number has been estimated at 1.4\%, but exact leakage rates are hard to determine (Lelieveld \textit{et al}. 2005). Leaking upwards of 25\% natural gas is achievable in a short timeframe if done intentionally, however. As demonstrated in Fig. 2a, and easily seen in Fig. 2b, leaking this additional amount of methane would be able to almost perfectly counteract the impacts of “moderate” geoengineering in a short timeframe. The radiative forcing of baseline RCP 4.5 and that of a world with both SRM and methane release start diverging more as “older” methane starts to decay, but the difference remains very small in the first 5 years after the start of counter-geoengineering. SRM’s radiative forcing is nearly 10 times higher after 5 years in a world without counter-geoengineering, suggesting the feasibility of methane as a way to derail geoengineering efforts in the short term.

\begin{figure}
\centering
    \includegraphics[width=1\linewidth]{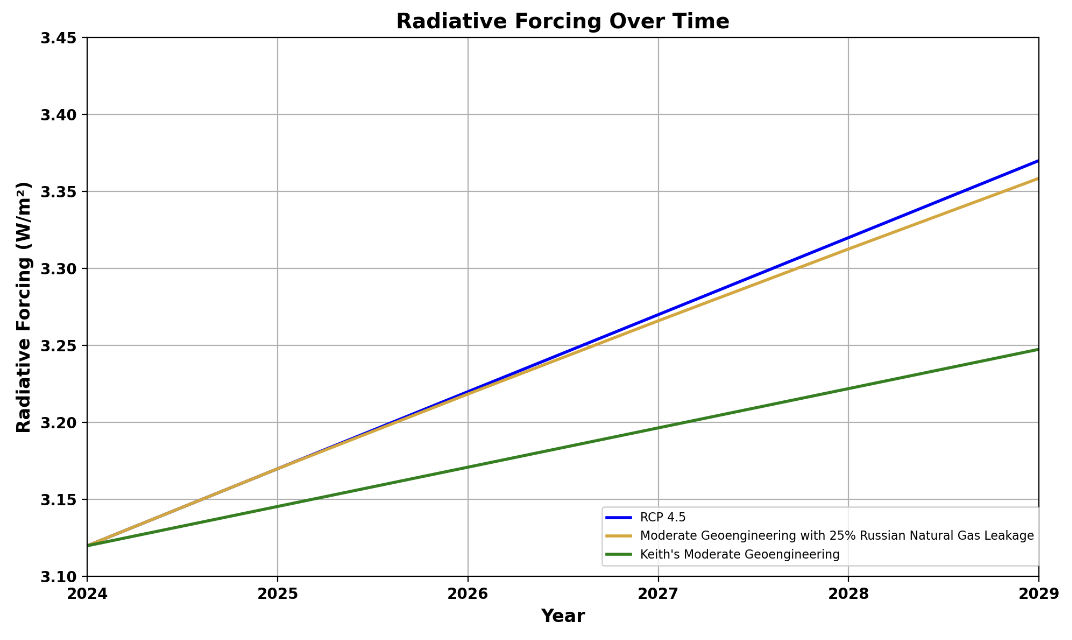}

    \includegraphics[width=1\linewidth]{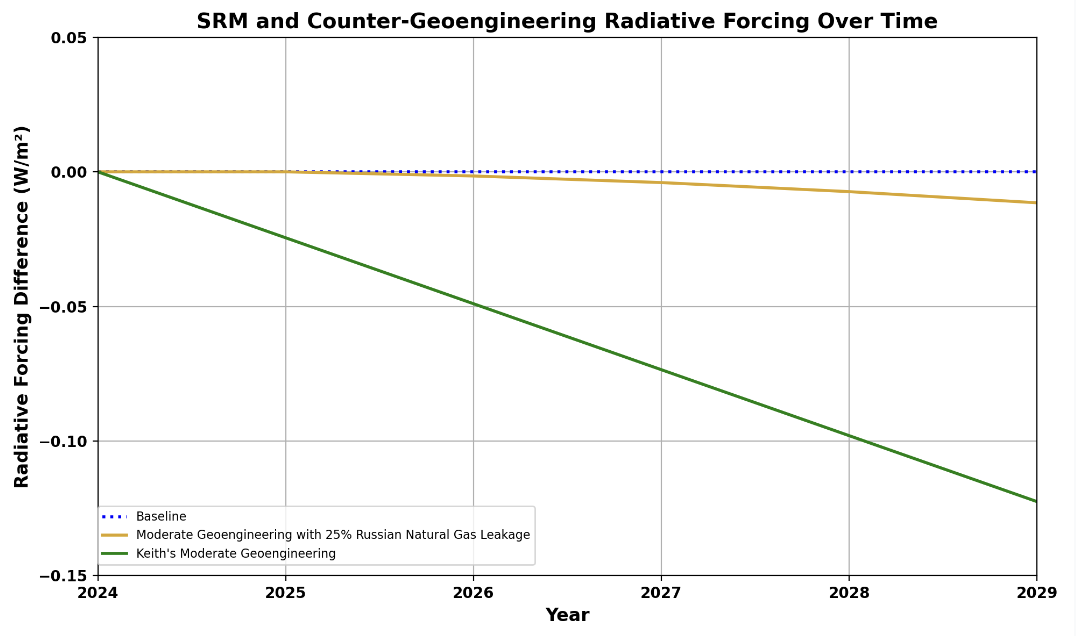}

    \caption{{(a) }This figure shows the overall net radiative forcing over a 5-year period under three different scenarios. The blue line portrays the baseline case of an RCP 4.5 scenario. The green line represents a scenario with a “moderate” geoengineering solution based on Keith and MacMartin (2015)’s solution of halving the rate of change of radiative forcing. Finally, the yellow line takes this moderate geoengineering solution and adds an additional 25\% leakage in natural gas from Russia. \textbf{(b) }This figure shows the radiative forcing of additional human interventions over a 5 year period, compared to a baseline scenario. The dotted blue line represents the base case of RCP 4.5, where no geoengineering or deliberate methane leakage is undertaken. The green line, as in 2a, represents a scenario based on moderate geoengineering, where each year the magnitude of SRM’s radiative forcing increases by 0.0245 W/m\textsuperscript{2}. The yellow line represents the radiative forcing of moderate SRM in conjunction with Russian methane counter-geoengineering in the form of an additional 25\% leakage in natural gas.}
\end{figure}
\clearpage

\section*{Results}

This moderate geoengineering solution has the intention to halve the growth rate of human-caused temperature change. In Fig. 2a, a geoengineered world would be expected to have a total human-caused radiative forcing of around 3.25 W/m\textsuperscript{2} after 5 years from the beginning of geoengineering implementation. In a world where nothing additional is done to combat climate change, simulated as the IPCC’s RCP 4.5 scenario, human-caused radiative forcing would be around 0.12 W/m\textsuperscript{2} higher in five years, at 3.37 W/m\textsuperscript{2}. Thus, geoengineering represents a significant mitigation of human-caused temperature change even within a short 5-year timeframe, with the difference between scenarios becoming greater the longer geoengineering is in place. 

The inclusion of counter-geoengineering, however, virtually eliminates the difference between RCP 4.5 and a scenario with geoengineering. By leaking an additional 25\% of its natural gas production, Russia could bring human-caused radiative forcing to 3.36 W/m\textsuperscript{2} after 5 years, roughly 0.01 W/m\textsuperscript{2} less than a scenario with no geoengineering. Naturally, the same result would follow if any other country, or combination of countries, released an additional 0.12 Gt of methane per year. Figure 2 gives a good visualization of the drastic reduction in the effects of moderate geoengineering in a world where an additional 0.12 Gt of methane is released into the atmosphere each year. In the first couple of years, methane release is able to nearly eliminate SRM’s radiative forcing. After 5 years, even as methane’s warming potential starts to decline, additional leakage is still able to reduce the effects of SRM by around 90\%. 

\section*{Discussion}

As the yellow line’s departure from RCP 4.5 in Figure 2 suggests, in longer timeframes it is likely that geoengineering would eventually have a significant effect as the warming potential of methane declines over time. It may also be increasingly costly to maintain a high level of methane leakage. However, within 5 years it seems feasible for a country to eliminate the effects of geoengineering by deliberately increasing its methane leakage from natural gas production, which can have consequential effects for the initial rollout of geoengineering. An additional 25\% leakage in natural gas production would significantly reduce Russia’s earnings from natural gas, which may be costly as time goes on. However, in the short-term, drastic reductions in natural gas production would likely be feasible. Between 2020 and 2022, Russia’s natural gas production fell by over 16\%, and other areas of the world have seen larger falls in a similarly short period of time (Enerdata 2024). Although a 25\% fall in natural gas production would be very significant, it may not pose enough of a cost in the short term to deter a counter-geoengineering response. Additionally, this calculation relies solely on methane, and likely overestimates the required amount of natural gas leakage to counteract SRM’s radiative forcing as natural gas’ non-methane components contain other warming agents. Other previously mentioned warming agents, such as chlorofluorocarbons, are dozens of times more potent than methane and could also aid in counter-geoengineering. Although they have much higher lifetimes than methane and are thus considered implausible as counter-geoengineering solutions, a very small release could have a very significant effect. Finally, we are conservative in our use of the higher end of SRM’s estimated radiative forcing, further indicating that counter-geoengineering may be realizable in the short term. 

The viability of counter-geoengineering as a way to keep the world on its current warming trajectory in a “moderately” geoengineered world makes geoengineering an infeasible solution to climate change. More aggressive versions of geoengineering have already faced significant pushback due to their potential risks, and have consistently been rejected by global leaders (Civillini 2024, Limb 2024, and Tollefson 2024). Furthermore, large-scale geoengineering would likely provide even more incentive for aggressive response from countries like Russia due to a possible self-defense argument resting on a quantifiable increase in cold-related deaths. This paper shows that solutions deemed moderate, which may pose less risks, would have to contend with their own set of problems. It is all but certain that global adherence to the deployment of geoengineering would not be complete. Not only is this already suggested by the contentious nature of the solution in academic circles, but it is also supported by the fact that some countries stand to benefit economically and geopolitically from rising temperatures. The existence of a viable way to counteract moderate geoengineering poses an existential threat to the solution’s achievability, as any effort to maintain the solution’s effectiveness in a counter-geoengineered world would create great levels of geopolitical tension. 

In a world that lacks complete agreement on the deployment of geoengineering, it is virtually impossible to impose the solution on every country. This is especially true given that geoengineering with methane release is unpredictable and can be deployed at a moment’s notice without warning. Countries implementing geoengineering would have to contend with governability issues inherent to solutions implemented on a planetary scale with worldwide consequences. In the case of geoengineering, concerns over the governability of its implementation have even prompted calls for an international non-use agreement (Biermann \textit{et al}. 2022). The possibility of conflicts arising from geoengineering is a major threat to the solution’s feasibility. These conflicts may pose significant security issues globally, with the risk of a collapse in global cooperation, and even the potential of war (Corry 2017). Conflicts may be even more likely with successful counter-geoengineering, as tensions rise between countries on different sides of the issue. Countries implementing a moderate version of geoengineering would have no plausible way of interrupting counter-geoengineering by other countries without a large global conflict, thus rendering their solution ineffective. An alternative to forceful adherence to geoengineering would be a ramp-up in stratospheric aerosol release, though, as widely discussed, greater levels of geoengineering pose much greater ecological and security risks that may not be worth taking. 

This analysis lends itself to a game theory approach. If we think of ourselves, under current mitigation measures, as in an equilibrium, geoengineering would represent a deviation that is at first positive for countries responsible for the solution and negative for others. Counter-geoengineering would then represent a retaliatory deviation that would benefit some countries as compared to a geoengineered world. This would lead the world into a new equilibrium, where both sets of countries are worse off than they were initially. The temperature remains the same due to the counteracting effects of sulfur and methane, but the world now relies on sulfur and methane release to keep that equilibrium. If an agreement is reached to stop both counter-geoengineering and geoengineering, there would be a much faster increase in worldwide temperatures than current rates as a result of the excess methane in the atmosphere that aerosols suddenly would not counteract. This makes for an even worse end-state scenario for geoengineering parties, as they would have to incur the costs of geoengineering with no way to make an agreement to leave the worse equilibrium state. 

A moderate geoengineering solution would potentially leave the world on the same warming trajectory that it is currently on, but with greater levels of geopolitical tension and much greater amounts of methane in the atmosphere from increased leakage rates. The feasibility of counter-geoengineering in a short period of time, as demonstrated by this paper’s finding that only an additional 0.12 Gt of methane would have to be released per year into the atmosphere, suggests that enacting a moderate geoengineering solution might cause the world to be in a more precarious place than it currently is. Counter-geoengineering thus presents a considerable challenge to solar geoengineering and constitutes a grave threat to the geopolitical state of the world as well as to the global climate. The lack of a global governance system capable of eliminating the threat of counter-geoengineering drives the potential cost of implementing geoengineering to levels too high for the solution to be deemed feasible.

This work adds to previous work by Heyen \textit{et al}. (2019) and Parker \textit{et al. }(2018) by providing a specific and feasible way to implement counter-geoengineering in a short timescale. It also discusses the possibility of counter-geoengineering in the current geopolitical context, providing a logical basis for its implementation and discussing its potential consequences. There are many ways in which future papers can improve or add to the results of this paper. Firstly, the calculations included in this paper only consider the effect of methane in analyzing the warming effect of natural gas leakage. A calculation that also includes the impact of other warming agents in natural gas may arrive at a more accurate, and likely smaller, leakage rate. Future research may also expand on this paper by including more detailed modeling of the effects of methane release, which may account for spatial considerations as well as different rates of decay in the atmosphere. Additionally, although the additional 0.12 Gt of methane per year would counteract the moderate version of geoengineering explored in this paper, more aggressive versions of geoengineering would require a greater response, which may not currently be feasible simply through methane leakage. We argue that a version of geoengineering that is impractical to counteract with methane release is not feasible due to the potential risks associated with great levels of stratospheric aerosol release. However, future research or technologies may alleviate concerns about greater aerosol release, in which case other counter-geoengineering alternatives would need to be researched.

\section*{Conclusion}

Geoengineering, even in its moderate forms, faces significant challenges that risk undermining its feasibility as a viable solution to combat climate change. While geoengineering may at first glance appear to offer a low-cost method to offset global temperature rise, the potential for counter-geoengineering against ‘moderate’ versions of geoengineering present a profound obstacle. Three scenarios exist, and they are either unrealistic or unattractive. First, a global agreement on geoengineering with an enforcement or oversight agency and based on a global consensus might preempt efforts to ‘counter-geoengineer.’ Such an agreement does not exist at present and is unlikely to be feasible in the near or medium term. Second, counter-geoengineering could be met with the use or threat of force by countries engaged in ‘moderate’ geoengineering. Because counter-geoengineering is relatively effortless, this scenario is likely to inflame geopolitical tensions and conflicts, potentially destabilizing international relations and leading to severe security threats. Large amounts of methane could be deployed at a moment’s notice without much difficulty and with virtually no warning to counteract SRM. This could at worst be catastrophic. Three, the absence of a global agreement where counter-engineering is not met with force or the threat of force could instead lead to greatly increased geoengineering, far past the ‘moderate’ geoengineering that is being proposed. This paper contributes to the ongoing discourse by providing a specific example of how counter-geoengineering could be implemented, reinforcing the argument that any ‘moderate’ geoengineering effort lacking comprehensive global support is likely to fail. As geoengineering is still touted by many as a low-cost and highly effective short-term fix, it is crucial that potential implications, harmful countering responses, and its lack of enforceability be considered.

\end{document}